\documentclass[conference]{IEEEtran}
\IEEEoverridecommandlockouts
\usepackage{cite}
\usepackage{amsmath,amssymb,amsfonts}
\usepackage{algorithmic}
\usepackage{graphicx}
\usepackage{textcomp}
\usepackage{xcolor}
\usepackage{csquotes}

\def\BibTeX{{\rm B\kern-.05em{\sc i\kern-.025em b}\kern-.08em
    T\kern-.1667em\lower.7ex\hbox{E}\kern-.125emX}}
\begin{document}

\title{LSTFCoDel: CoDel with LSTF-Style Priority Queuing\\}

\author{\IEEEauthorblockN{Christen Ford}
\IEEEauthorblockA{\textit{Electrical Engineering and Computer Science} \\
\textit{Cleveland State University}\\
c.t.ford@vikes.csuohio.edu}\\
\textit{IEEE Student Member}
}

\maketitle

\begin{abstract}
Congestion control is vastly important in computer networks. Arising naturally from the bursty nature of Internet traffic, congestion plagues not only the network edge, but also the network core. Many remedies have been proposed to fight congestion; active queue management (AQM) is one such proposal. AQM seeks to prevent congestion by actively avoiding it. 

Some queuing disciplines such as Random Early Detection (RED) will prematurely drop a random packet (with some probability) when the queue nears capacity to signal the sender to back off. However, RED utilizes queue length as a mechanism to indicate congestion. On the other hand, the Controlled Delay (CoDel) queuing discipline uses queuing delay as an indication of congestion.

The problem with both RED and CoDel are that they indiscriminately treat all packets the same. Normally implemented using a FIFO queue, CoDel simply enqueues and dequeues packets in a first-come, first-served manner. Priority queuing can be carefully utilized to selectively service packets utilizing the very same metric CoDel uses for AQM, queuing delay. That said, Least Slack Time First (LSTF), a multi-processor scheduling algorithm employs priority scheduling, which coincidentally, is also based on delay.

In the context of computer networks LSTF can be applied in the control plane or in the data plane. At the control plane, LSTF functions across the entire network, but in doing so requires all intermediary routers to implement it; LSTF also requires support at the packet level in terms of a slack entry. Within the data plane, LSTF can be implemented as a queuing mechanism based on delay spent in the router (just like CoDel AQM). This paper applies data plane level LSTF to CoDel AQM to enable delay-based packet classification within the confines of the CoDel AQM algorithm.
\end{abstract}


\section{introduction}
Bufferbloat is a major contributor to congestion in the Internets network core. While high-traffic links and expensive switching devices can help with bufferbloat, the heart of the issue lies with the queuing algorithms employed on switching devices. Much time and money has been spent finding new ways to fight bufferbloat, however, the key to fighting bufferbloat lies with cooperation with the transport layer in end-hosts and the network and data link layers in intermediary switching devices. 

An interesting mechanism (not necessarily geared towards congestion control or avoidance) requiring support both at the transport and network layers is Least Slack Time First (LSTF). LSTF is implemented network wide and provides prioritized packet service based on the time needed for a packet to reach a destination host from the sending host. This time is known as 'slack' and is accounted for at each intermediary router, where packets with the least slack time receive priority service. This mechanism can be adapted for use with congestion avoidance mechanisms to provide priority service to packets that would otherwise be unfairly dropped due to congestion.

The Transmission Control Protocol (TCP) is one of two primary protocols used within the transport layer of the modern Internet architecture. Introduced in \cite{tcp_paper}, TCP provides all of the functionality that the User Datagram Protocol (UDP) does not. Specifically TCP provides reliable, in-order delivery of packets at the expense of increased packet overhead, and implementation complexity. TCP also provides congestion control through the use of observed feedback of the network as well as explicit feedback from intermediary switching devices and the destination host. One of the mechanisms by which TCP does this is through round-trip time estimation. This estimation is tied to the TCP send window and is used to help prevent the protocol from overwhelming the network and the destination host.

Explicit Congestion Notification (ECN) is an extension to the network layer Internet Protocol (IP) that allows ECN aware routers to provide congestion avoidance. ECN has a simple implementation but requires the modification of IP packet headers to enable the use of an ECN bit. When a router supporting ECN receives a packet and detects congestion, rather than dropping the packet, it will set the ECN bit and send a response back to the sender. The packet is forwarded to the next router (which hopefully is also ECN aware) which is then made aware that it should expect congestion. Meanwhile, the sender will reduce its sending rate upon receiving the congestion notification from the router that first set the ECN bit. An alternative to ECN that does not depend on multiple routers implementing ECN functionality is Random Early Detection (RED).

RED is an AQM queuing discipline that utilizes queue length as a means to gauge congestion. RED operates in three stages determined via a window on queue length using a minimum and maximum threshold. In the first stage, RED operates like any other queuing discipline, packets arrive and eventually receive service. The second stage is reached when the queue length meets the minimum threshold, packets are marked with some sliding probability that is dependent on queue length. Finally, once the queue length meets the maximum threshold, RED enters into the third stage. In this stage, RED will always mark at least one packet to indicate congestion. This stage is reached when the queue length meets some maximum threshold. RED routers (or gateways as they are called in \cite{red_paper} are an effective AQM mechanism, however, queue length is not always a reliable measure of congestion, as will be discussed.

Building on top of ECN and RED, Nichols, Jacobson, McGregor, and Iyengar  introduce the Controlled Delay (CoDel) AQM packet scheduling algorithm \cite{codel_rfc} to help fight bufferbloat. CoDel narrows the scope of controlling delay to within the switching device itself. CoDel lets so-called "sojourn time" (the time a packet spends in the switching device) indicate whether congestion is occurring or not i.e. whether the current queue is a bad queue or a good queue. CoDel serves as one of the two primary components of LSTFCoDel, the other being Least Slack Time First.

Least Slack Time First (LSTF), originally conceived in \cite{lstf_paper}, is a multi-processor scheduling algorithm for jointly scheduling tasks on systems with multiple physical processors. The major idea behind LSTF is so-called "slack time". Slack time is employed by the algorithm in a priority fashion to determine which tasks execute in what order. When a process first enters the system, it is assigned an amount of estimated slack time. This slack time is then decreased by the amount of time the process spends as the running process on each processor. This mechanism is easily portable to the realm of computer networks, and can be used to build fair local area networks as shown in \cite{unisched_paper}. Within the context of LSTFCoDel, I employ it as the main driver for prioritizing CoDel. However, the determination of what slack means within a single router, comes from TCP.

\section{Related Works}
\subsection{Transmission Control Protocol}
The transmission control protocol (TCP) is highly influential in the design of LSTFCoDel. TCP is used by end hosts at the transport layer for reliable, in-order transmission and receival of packets. This will not be a thorough discussion of TCP, readers interested in the protocol are directed to \cite{tcp_paper}. For the interests of this paper, we are specifically interested in classical TCP RTT estimation as defined in \cite{karn_partridge_1991}. Given an $\alpha$ value in the range of [0, 1]. The formula for estimating RTT by ``Eq.~\eqref{eq:tcprtt}''.

\begin{equation}
    \gamma = (1-\alpha) * \gamma + \alpha * \delta
\label{eq:tcprtt}
\end{equation}
where $\gamma$ and $\delta$ are the estimated RTT and sample RTT respectively.

The strength in this formula is that it considers both historic RTT as well as sampled RTT. This formula has been thoroughly researched since the inception of TCP and its use in computer networking is highly proliferate. This formula is employed by LSTFCoDel in estimating the average slack time per packet.

The application of this formula to the problem of estimating switching device level slack time provides several advantages:
\begin{itemize}
    \item Slack time is only dependent on whats going on at the device itself. Slack time is completely independent of whats going on with the rest of the network.
    \item This formula allows slack time to take into consideration current delay as well as historical delay experienced at the device.
    \item No proof of the success of this formula is necessary, its success is already proven through its use in TCP as well as various other networking protocols and applications.
\end{itemize}

ECN is a transmission control protocol (TCP) modification that marks packets with an ECN flag. ECN allows networks to respond to congestion without needing to explicitly drop packets. With ECN, it is the routers responsibility to decide whether a packet should have its ECN flag set or not. According to her experiments with ECN in \cite{tcpecn_paper}, Sally Floyd states that "For TCP, the receipt of a single ECN (e.g., a single Source Quench packet, or a single packet with the ECN bit set) should trigger a response to congestion." The switching device passes ECN packets along just like non-ECN packets. The exception being that when congestion occurs, ECN packets are not dropped by the router, they are intentionally forwarded to the receiver to signal that congestion is imminent.

\subsection{Least Slack Time First}
Least Slack Time First (LSTF) introduced in \cite{lstf_paper}, is a priority scheduling discipline that has been shown in \cite{unisched_paper} to viably replay many different scheduling algorithms. As described in \cite{lstf_paper}, LSTF was originally indented to provide scheduling for multi-processor systems. The reader is urged to make the distinction that multi-processor system refers to a system with multiple physical processors not a single processor with multiple logical cores. 

As stated in \cite{unisched_paper}, we can adopt an approach for slack in computer networks where we define slack as either an observed or estimated transmission time between the source host and the intended receiving host. This slack time is then appended to the outgoing packets header before departure from the source host. As the packet works its way through the network, each router decrements the slack value by the queuing delay. In this way, it is ensured that packets with the Least Slack Time First receive service. This sense works fine in constrained environments such as LAN environments where the network can be more easily controlled, however, for larger networks this brings into discussion two fundamental issues:

\begin{enumerate}
    \item Packets must carry the slack time for LSTF to function.
    \item Intermediate devices in the route from sender to receiver must implement the LSTF algorithm.
\end{enumerate}
It should be inferred that the receiver does not need to implement LSTF for the system to function.

This definition of network wide slack time comes from \cite{unisched_paper}. It is an excellent scheme for smaller, more controlled networks, however, as shown, it has little practical use in real-world computer networks due to the two aforementioned problems. If a micro approach is taken with regard to slack time, then both problems become defunct. That is, implement LSTF not in the control plane (creating dependency issues in the process), but in the data plane. With this notion, slack exists within the confines of switching devices that need not be dependent on packets carrying slack time across the network or other routers implementing the LSTF scheduling algorithm.

There are several alternative metrics that could constitute slack time within switching devices, but the most intuitive is to allow the queuing delay to stand as slack time. However, due to congestion and other factors that affect queuing delay, a static assignment of slack time based on observed queuing delay is incorrect. In this paper, I present a method for determining slack time that is derived from estimating queuing delay using both observed delay as well as past delays experienced at the switching device. 

\subsection{Random Early Detection (RED)}
Random Early Detection (RED) is a queuing discipline first introduced by Sally Floyd and Van Jacobson in \cite{red_paper}. Routers that implement the RED queuing discipline are known colloquially as 'RED gateways'. RED is a form of AQM that maintains a sliding window of allowable congestion, based on queue length. The lower bound of this window indicates the point where RED will start marking packets with some probability. The upper bound indicates the point where RED will transition into dropping state. Once crossed, RED will mark the next packet received with 100\%  probability to immediately indicate congestion.

Throughout its operation, RED maintains an average queue size, which according to Floyd and Jacobson in \cite{red_paper}, "The RED congestion control mechanisms monitor the average queue size for each output queue, and, using randomization, choose connections to notify of that congestion. Transient congestion is accommodated by a temporary increase in the queue. Longer-lived congestion is reflected by an increase in the computed average queue size, and results in randomized feedback to some of the connections to decrease their windows." If the average queue size falls within some sliding window, then a packet is marked to be dropped with a certain probability (to be discussed later). If the average queue size falls outside the upper bound of the window, then the next received packet is always marked.

A theoretical explanation of RED is not necessary, however, it is worth it to mention how RED updates its average queue size as it is similar to the mechanism used to update estimated RTT in TCP. As described by Floyd and Jacobson in \cite{red_paper}, upon packet arrival, the average queue size is calculated as:
\begin{algorithmic}
\IF {the queue is nonempty}
    \STATE $avg\gets (1-w_q)*avg+w_q*q$ 
\ELSE
    \STATE $m\gets f(time-q\_ time)$
    \STATE $avg\gets (1-w_q)^m*avg$
\ENDIF
\end{algorithmic} 
Where \textit{avg} is the average queue length, \textit{q} is the current queue size, \textit{w$_\text{q}$} is a fixed queue weight, \textit{time} is the current time, \textit{q\_time} is the time the queue entered idle state, and \textit{f$($t$)$} is a linear time function.

Notably, we see that if the queue is not empty, RED will update the average queue length to some percentage of the historical queue length plus some percentage of the current observed queue length. This equation is identical to that used to update estimated RTT for classical TCP. It is also worth mentioning that the minimum and maximum thresholds used to determine the sliding marking window for RED are determined from the average queue length.

Additionally, Floyd and Jacobson point out that the proper selection of \textit{w$_\text{q}$} is paramount to the success of RED in detecting and controlling congestion. If \textit{w$_\text{q}$} is too low, then \textit{avg} will respond too slowly to changes in the queue size. Conversely, if \textit{w$_\text{q}$} is too high, then RED will fail to detect and filter temporary congestion in the gateway. Floyd also provides a derivation for both of these bounds, but the discussion of such are outside the realm of this paper. Interested readers are deferred to \cite{red_paper}.

\subsection{Controlled-Delay AQM}
The Controlled-Delay (CoDel) active queue management (AQM) algorithm formally defined in RFC 8289 \cite{codel_rfc} by Nichols et al. is a congestion control focused queuing discipline designed to fight the persistently full buffer problem (bufferbloat). Bufferbloat is a serious problem in today's computer networks that has only worsened with the prevalence of software-as-a-service (SaaS) and streaming websites. Irregardless of bandwidth, if congestion is left unchecked, delay will grow unbounded causing bandwidth to become irrelevant.

According to \cite{codel_rfc}, CoDel was designed with five goals in mind:
\begin{itemize}
    \item Make AQM parameterless for normal operation, with no knobs for operators, users, or implementers to adjust.
    \item Be able to distinguish "good" queue from "bad" queue and treat them differently, that is, keep delay low while permitting necessary bursts of traffic.
    \item Control delay while insensitive (or nearly so) to round-trip delays, link rates, and traffic loads; this goal is to "do no harm" to network traffic while controlling delay.
    \item Adapt to dynamically changing link rates with no negative impact on utilization.
    \item Allow simple and efficient implementation (can easily span the spectrum from low-end access points and home routers up to high-end router hardware).
\end{itemize}

So how does CoDel actually control delay without depending on parameters outside of the routers control? This is done purely from the context of packet-in, packet-out, that is, control of delay is handled exclusively through queuing delay (known as \textit{sojourn time} within the confines of CoDel). Additionally, CoDel employs a set-point as well as a state-space controller for maintaining state of the algorithm.

Under normal operation, CoDel acts very much like a FIFO queue (in fact, CoDel is a FIFO queue). Packets enter the switching device, are marked with an entry time, and are sent into the devices statistical multiplexer. Packets are serviced in the order they arrived. It is worth describing CoDel's dequeuing operation, as that is where the heart of CoDel lies.

During the dequeue operation, CoDel's state-space controller must differentiate whether the queue is in a "good" state or in a "bad" state. Remember the sojourn time and the set-point mentioned earlier? These two factors are employed to determine if the queue is in "good" state or "bad" state. Queue congestion is observed from the sojourn time. The reason sojourn time makes a good metric for congestion is rather intuitive. When packets enter the queue they are time-stamped. That time-stamp is then compared to a time-stamp of when the packet is dequeued. The difference indicates the delay the packet experienced in queue (the sojourn time). As congestion increases, the delay per packet will reflexively increase in turn, naturally indicating congestion.

Once the observed sojourn time becomes larger than the target set-point mentioned earlier, CoDel enters into a dropping state via an estimator function. In dropping state, CoDel's controller will drop packets until it learns an appropriate response to the traffic generating the congestion. This learned response is stored in what CoDel calls the "next drop time". The next drop time is according to \cite{codel_rfc}, "... [a] stochastic gradient learning procedure [that] is the core of CoDel's control loop (the gradient exists because a drop always reduces the (instantaneous) queue, so an increasing drop rate always moves the system "down" toward no persistent queue, regardless of traffic mix)." CoDel's next drop time increases slowly to prevent over-dropping and guarantee dissipation of the persistent queue. Conversely, the next drop time "is decreased in inverse proportion to the square root of the number of drops since the drop state was entered..." This relationship allows CoDel to service packets in FIFO fashion with normal congestion, while still allowing for bursty traffic.

It should be noted that CoDel's estimator function is not without error. According to \cite{codel_rfc}, when CoDel initially drops a packet due to the sojourn time crossing the target, CoDel will over-drop if the RTT (interval) and the target value are a longer time period then when the next drop actually happens. The cause of this is inherent congestion delays within the router. In practice, CoDel handles this by setting the drop times spacing to the estimator functions interval. 

CoDel has been very successful in the field. It was originally implemented in the Linux kernel back in 2012 and has spawned many derivatives including \cite{fqcodel_rfc} and \cite{sfqcodel_src}. The most notable of which is Stochastic Flow Queuing Codel (sfqCoDel) which employs flow queuing with deficit round robin (DRR) over a number of CoDel bins to appropriately service packets based on flow id. Packets enter the system and are hashed into a configurable number of bins based on several pieces of information in the packet header including the flow id. Due to the newness of the algorithm implementations of sfqCoDel in switching devices are rare, although thorough research exists backing sfqCoDel within the Network Simulator 2 (NS-2) environment.

\section{Methodology}
LSTFCoDel as the name implies combines techniques from the LSTF and CoDel queuing disciplines. From LSTF, LSTFCoDel derives the mechanism used to enqueue and dequeue packets. In terms of active queue management, LSTFCoDel and CoDel function identically, as such, I refer the reader to the CoDel discussion in the related works section. The major difference between LSTFCoDel and CoDel is the way the queue handles enqueuing and dequeuing under normal conditions.

As mentioned prior, LSTF maintains a slack time entry within the packet header to service packets with the least slack time first. It has already been shown that LSTF works extremely well in theory and simulation \cite{lstf_paper} and \cite{unisched_paper}, however, on large networks, LSTF quickly becomes impractical. For LSTF to properly function, each intermediate router must implement the LSTF algorithm. This is paramount, as the slack value is set to the estimated time from the source to the receiver at the ingress of the network. Each intermediate router updates the slack value of a received packet upon egress from the router to its prior value minus the time spent in the router. The backing queue services packets in a priority manner with the packets with the least slack time being serviced first.

LSTFCoDel allows the notion of slack time to be adhered to (in a slightly altered fashion) without:
\begin{enumerate}
    \item Requiring every router in the route implement LSTF.
    and
    \item Modifying packet headers to accommodate slack time.
\end{enumerate}
Before dealing with LSTFCoDel in depth, it is imperative to become familiar with the symbology that represents important parameters of the LSTFCoDel AQM algorithm. These symbols are listed in ``Table~\ref{table:lstfparams}''.

\begin{table}[htbp]
\caption{LSTFCoDel Parameters}
\begin{center}
\begin{tabular}{|c|c|}
\hline
     Symbol & Meaning \\
     \hline
     
     \hline
     $\alpha$ & LSTFCoDel's forgetfulness factor \\
     \hline
     $\beta$ & Current Delay Experienced \\
     \hline
     $\gamma$ & LSTFCoDel's average slack \\
     \hline
     $\epsilon$ & LSTFCoDel's packet priority \\
     \hline
\end{tabular}
\label{table:lstfparams}
\end{center}
\end{table}

LSTFCoDel is a priority queuing mechanism featuring CoDel AQM. Packets are classified upon ingress into the router using the average slack value as given by ``Eq.~\ref{slack_calc}''.

\begin{equation}
    \gamma = (1 \ \text{--} \ \alpha) * \gamma + \alpha * \beta
\label{slack_calc}
\end{equation}

Knowing $\gamma$ allows us to identify LSTFCoDels classifier function. The classifier function is given by ``Eq.~\ref{classify_func}''.
\begin{equation}
 \epsilon = 
 \begin{cases}
      0 & \gamma = 0 \\
      \frac{1}{1 + \gamma} & otherwise
   \end{cases}
   \label{classify_func}
\end{equation}

Understandably, $\gamma$ has the potential to be zero but it will never be less than zero. The classifier function uses an inverse of $\gamma$ because we want to give preference to packets that become delayed due to temporary congestion. Additionally, one is added to $\gamma$ to ensure that the priority is calculated correctly. Assuming no excessive delays are present, $\gamma$ will tend to fall in the range [0, 1]. Under normal conditions i.e. no congestion, LSTFCoDel will continue to function normally, largely servicing packets in the order they arrive.

$\beta$

$\alpha$ represents what I have dubbed as LSTFCoDel's forgetfulness factor. It is employed by LSTFCoDel to determine the influence current slack time and the next estimated drop time have on the new slack time. If everything is fine with the router and no congestion occurs (implying CoDel has not been activated), then the estimated drop time will persistently be 0. After CoDel's \textit{drop next} parameter is utilized to update the slack value, it is then reset to 0 to prevent it from erroneously influencing the next slack update. This has no effect on the performance of CoDel, as CoDel does not depend on past values of its \textit{drop next} parameter for future dequeue decisions.

$\gamma$ is LSTFCoDel's average slack value. The slack seen here is different from the slack value seen in [reference to leung here]. The slack defined in the aforementioned paper is determined at the ingress of the packet into the network and decremented at each intermediate router until the packet reaches the destination host. This is in contrast to the slack defined by LSTFCoDel. Here, we let slack be a combination of the expected and observed delay packets face in the queue. The most important feature of slack in this definition is that slack is adaptive. Slack adapts to congestion so it functions in concert with CoDel.

$\epsilon$ is LSTFCoDels calculated priority value. This value is determined when a packet enqueues and determines the packets dequeue order. Because $\epsilon$ is calculated using delay history as well as current delay, it slides around based on congestion level. $\epsilon$ is heavily influenced by $\alpha$. An alpha value that is too low or too high incurs issues with the dequeing operation. Part of this research involves determining an initial best value for $\epsilon$ from intense simulation and statistical analysis. If $\epsilon$ is accurate of actual delay, then LSTFCoDel will dequeue otherwise heavily delayed packets due to congestion. 

With a thorough understanding of the inner workings of LSTFCoDel, I can now provide the motivation behind its development. As stated, under normal, non-congestion conditions, CoDel just like a FIFO queue while LSTFCoDel approximates a FIFO queue. The reason LSTFCoDel approximates a FIFO queue is because of subtle variations in delay experienced both at and on the way to the switching device. These delays will slightly alter the priority assigned to each packet. The only way LSTFCoDel will function as a FIFO queue is if each packet is assigned exactly the same priority upon entering the switching device.

When congestion occurs, CoDel and LSTFCoDel will function exactly in the manner described in the subsection on CoDel in the related works section above. However, since CoDel is merely a FIFO queue, all packets will be treated the same regardless, meaning packets that enqueue when the switching device is already suffering from congestion are likely to be dropped. This is unavoidable really, partially because of the way CoDels AQM functions, but also because there is no more buffer space available. Packets that survive CoDels dropping state are on average much more likely to experience longer queuing delay than those that were already in the front of the queue (because of the FIFO principle at work).

This is undesirable; should packets be delayed more because of network facets outside of their control? This does not have to be the case. LSTFCoDel employs a priority queue, where each packets priority is based on the average delay experience by packets in the system. Under congestion, arriving packets will be assigned a priority value that is much lower than those already in the queue. When congestion settles down, the average queuing delay will become more stable, therefore the priorities assigned to incoming packets will be more tightly grouped.

By weighting packet departure, LSTFCoDel ensures that otherwise heavily delayed packets receive service ahead of time. This ensures that the average queuing delay does not grow excessively large (as would happen in CoDel due to the FIFO queue). FIFO queues always service packets in the order they arrive. If congestion occurs, then as a consequence packets that arrive during congestion will always be heavily delayed. This overall produces the net effect of LSTFCoDel having on average less queuing delay than CoDel, which in turn affects other network factors such as RTT.

\section{Data Gathering and Analysis}
Before discussing LSTFCodel methods and results, I would like to explain the base case, CoDel. As stated, the primary motivator of LSTFCoDel is to provide equitable service in times of congestion. Remember that CoDel (and most other AQM methods) indiscrimantely treat all packets the same when congestion occurs. That is they drop arriving packets. Should these packets be dropped? Is it fair to drop them? These are some of the questions I seek to answer with this research.

This will not be a platform for the politics of packet scheduling, merely, I aim to provide a scheduling mechanism with sliding equity. In times of congestion, prioritize packets that will be immediately dropped; without congestion, function normally. In between, the algorithm should selectively prioritize packets based on observed and historical queuing performance. Anyway, on to CoDel.

Both CoDel and LSTFCoDel were tested in identical conditions with an identical run time of 259,200 seconds or three days. That said, metrics on CoDel's simulation are given in ``Table~\ref{table:codel_delay_metrics}'':
\begin{table}[htbp]
\caption{CoDel Packet Delay Statistics}
\begin{center}
\begin{tabular}{|c|c|}
\hline
     Metric & Result \\
     \hline
     
     \hline
     Mean & 0.03532900 seconds \\
     \hline
     Variance & $5.2169810^{-06}$ seconds \\
     \hline
     Std. Dev. & 0.00228407 seconds \\
     \hline
\end{tabular}
\label{table:codel_delay_metrics}
\end{center}
\end{table}

Likewise, the queue length metrics for CoDel are given in ``Table~\ref{table:codel_qlen_metrics}''. Note that for both CoDel and LSTFCoDel the maximum transmission unit (MTU) of all links in the simulation is 1,500 bytes in correspondence with Ethernet. For CoDel, the results in the previously mentioned table indicate that CoDel had on average approximately $7593.77bytes / 1500bytes \approx 5\ packets$ in the queue. The variance and standard deviation show that CoDel experienced a wide range of queue length indicating varying levels of congestion.
\begin{table}[htbp]
\caption{CoDel Queue Length Statistics}
\begin{center}
\begin{tabular}{|c|c|}
\hline
     Metric & Result \\
     \hline
     
     \hline
     Mean & 7,593.77 bytes \\
     \hline
     Variance & 557,311 bytes \\
     \hline
     Std. Dev. & 746.532 bytes \\
     \hline
\end{tabular}
\label{table:codel_qlen_metrics}
\end{center}
\end{table}

Do note that these metrics (and those presented in the following sections) were calculated using the method described by J. Cook (and originally proposed by Knuth) in \cite{incr_var_web}. These values were computed across the entire population of the data set for the respective algorithm. In a future section, we will perform hypothesis testing using normalized distributions of this data that are computed from these metrics (via the Central Limit Theorem).

LSTFCoDel was implemented within the Network Simulator 2 (NS-2) network simulation environment, the source code for which is indicated in \cite{lstfcodel_src}. LSTFCoDel's implementation acts on its own without any need to modify other NS-2 classes. It is an entirely plug-n-play module (just like CoDel). There are a few quirks to this implementation that are worth mentioning here. They are largely due to my own issues working with the NS2 source code in relation to the lack of proper developer documentation within the environment itself. The major decisions made regarding LSTFCoDel's implementation are as follows: \\
\begin{itemize}
    \item LSTFCoDel's priority queue mechanism is implemented as an STL multimap. This restricts enqueuing to the run-time exhibited by multimap's insertion routine, however, because the multimap maintains ordering, dequeuing is always on the order of \textit{$$O(1)$$} since dequeuing always pulls the first element from the multimap. \\
    \item Because of the way queues are traditionally implemented in NS-2, the introduction of the multimap to maintain priority queuing leads to a shadowing of every packet in the system. This implementation carefully accounts for the shadowing, so no packets are left in either the queue or multimap. Just be aware that the space requirements for LSTFCoDel in this implementation are on the order of \textit{$$O(n^2)$$} where \textit{n} is the number of packets in the queue. \\
    \item LSTFCoDel utilizes NS-2's \textit{PacketQueue} class as the backing queue for buffering packets. \textit{PacketQueue} allows for arbitrary removal of packets based on the packet itself through the use its \textit{remove} method. If the packet is not found in the queue, the simulation is forced to abort. This method is used to support priority queuing as opposed to using the \textit{deque} method of \textit{PacketQueue}. As a consequence of sequentially searching for the packet to remove as opposed to removing the last packet, this functionality operates on the order of {\textit{$$O(n)$$}} where \textit{n} is the number of packets in the queue. This is an obvious trade-off for not implementing my own priority queue in NS-2 (although I did try), the impacts of which are noted in the drawbacks and future works section. \\
\end{itemize}
The simulation script may be found at the same site the NS-2 implementation is available from. The simulation is run seven times with the following values for $\alpha$: 0.125, 0.25, 0.375, 0.5, 0.625, 0.75, 0.875. As with CoDel, each simulation is run for a total of 259200 seconds or three days. The topology for the simulation is as follows:

\begin{figure}[htbp]
\centerline{\includegraphics[scale=0.3]{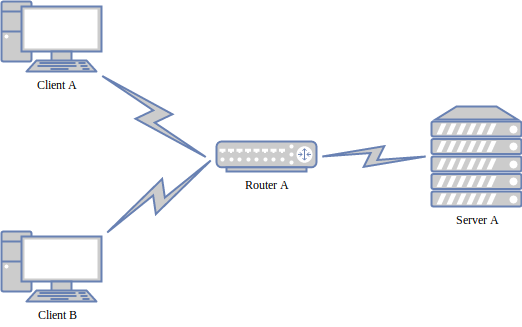}}
\caption{NS-2 Simulation Topology}
\label{fig:topology}
\end{figure}

The preceding topology consists of two client machines 'Client A' and 'Client B' and one server 'Server A' connected to the same router 'Router A'. Client A generates file transfer protocol (FTP) traffic over TCP and is connected to Router A over a 2Mbps link. Client B generates constant bit rate (CBR) traffic over the User Datagram Protocol (UDP) and is connected to Router A over a 1.5Mbps link. Finally, Server A is connected to Router A over a 1.7Mbps link. These parameters were carefully chosen to balance CoDel between dropping and non-dropping state. It should also be mentioned that both File Transfer Protocol (FTP) starts at $time = 0$ and Constant Bit Rate (CBR) traffic starts at $time = 300$ and run per the length of the simulation.

Prior to this simulation I attempted to design a simulation in NS-2 with N clients and N servers where each client would have a percentage chance to be a TCP client or a UDP client. Each client would be randomly connected to a server (so it was possible for some servers to have multiple clients, and some to have none). The simulation itself was configured for a number of transfer rounds where each client would be assigned a time window within each round to where it would transmit traffic to its assigned server.

I ended up abandoning this simulation for the following reasons:
\begin{itemize}
    \item As mentioned, there were some servers who received traffic from multiple sources, while some received traffic from none.
    \item The script was difficult to maintain as it was written in Tool Command Language (TCL) per NS-2 which becomes cumbersome with larger programs.
    \item The randomized nature of the script made it difficult to generate repeatable results.
    \item NS-2 finds it difficult to cope with many different events scheduled at roughly the same time. Most of the time, when running the script, NS-2 would error out because of a seemingly present timing error regarding same client transmission scheduling.
\end{itemize}

The simplified topology presented in this paper eliminates these issues and makes the resultant data-set far easier to analyze.

The following discussions provide statistical insight into both CoDel and LSTFCoDel. A few notes must be taken into consideration when digesting the following sections:
\begin{itemize}
    \item There is a single control for CoDel. In any statistical tests that follow, LSTFCoDel is compared solely against it.
    \item Statistical metrics are computed across the enitre population. Population sizes are not the same and are quite large. Using the Central Limit Theorem it is thus possible to generate normally distributed samples of which tests can then be performed against.
    \item All simulations, both CoDel and LSTFCoDel were run with the exact same parameters for the exact same amount of time with the exact same parameters for all clients, traffic generators, and the CoDel AQM algorithm.
\end{itemize}

\subsection{LSTFCoDel Queue Length, Delay, and Slack}

Queue length is normally used as both a metric for congestion as well as a means to determine the effectiveness of a queuing discipline. Good queuing disciplines should be able to handle bursty internet traffic while allowing for constant streaming traffic as is done with \cite{codel_rfc}. I provide queue length statistics for LSTFCoDel for all values of $\alpha$ tested in this paper in ``Table!\ref{table:lstfqlen}"". 

\begin{table}[htbp]
\caption{Queue Length Statistics for LSTFCoDel $\alpha$ Levels}
\begin{center}
\begin{tabular}{|c|c|c|c|}
\hline
     Forgetfulness ($\alpha$) & Avg. (bytes) & Var. (bytes) & Std. Dev. (bytes) \\
     \hline
     
     \hline
     0.125 & 1,598.87 & $4.10871*10^6$ & 2,026.99 \\
     \hline
     0.250 & 1,512.69 & $3.77956*10^6$ & 1,944.11 \\
     \hline
     0.375 & 1,754.56 & $6.82233*10^6$ & 2,611.96 \\
     \hline
     0.500 & 1,370.88 & $3.20691*10^6$ & 1,790.78 \\
     \hline
     0.625 & 1,722.86 & $5.21144*10^6$ & 2,282.86 \\
     \hline
     0.750 & 1,608.45 & $4.15869*10^6$ & 2,039.29 \\
     \hline
     0.875 & 1,682.16 & $3.46629*10^6$ & 1,861.80 \\
     \hline
\end{tabular}
\label{table:lstfqlen}
\end{center}
\end{table}

The results in ``Table~\ref{table:lstfqlen}"" are indicative of lower average queue length than CoDel, however, the variance and standard deviation are much higher than CoDel indicating queue length jumped up and down a lot more than CoDel. CoDel may display overall higher queue length than LSTFCoDel but at least as far as this particular implementation is concerned, LSTFCoDel is far less stable. The same story is replayed for delay over time.

Delay over time is an important metric to measure for LSTFCoDel. Recall that CoDel directly employs delay in its congestion avoidance mechanism; while LSTFCoDel uses delay to determine packet priority. In this section, I present a statistical analysis of delay over time as well as a comparison to the CoDel control simulation. ``Table~\ref{table:lstfdelay}'' shows standard statistical metrics for LSTFCoDel given all $\alpha$ values tested in this research. These metrics include the mean, variance, and standard deviation.

\begin{table}[htbp]
\caption{Delay Statistics for LSTFCoDel $\alpha$ Levels}
\begin{center}
\begin{tabular}{|c|c|c|c|}
\hline
     Forgetfulness ($\alpha$) & Avg. (s) & Var. (s) & Std. Dev. (s) \\
     \hline
     
     \hline
     0.125 & 0.01051590 & 0.000713601 & 0.0267133 \\
     \hline
     0.250 & 0.00984854 & 0.000736556 & 0.0271396 \\
     \hline
     0.375 & 0.01158550 & 0.001640950 & 0.0405087 \\
     \hline
     0.500 & 0.00859185 & 0.000328906 & 0.0181358 \\
     \hline
     0.625 & 0.01101150 & 0.000932091 & 0.0305302 \\
     \hline
     0.750 & 0.01051490 & 0.000859984 & 0.0293255 \\
     \hline
     0.875 & 0.01128840 & 0.000844949 & 0.0290680 \\
     \hline
\end{tabular}
\label{table:lstfdelay}
\end{center}
\end{table}

Several conclusions can be drawn from ``Table~\ref{table:lstfdelay}''. First, the average delay experienced with LSTFCoDel is always less than the average delay experienced within CoDel. Second, The variance or spread of delay is much higher with LSTFCoDel. Third, LSTFCoDel's standard deviation is higher than CoDel. These results and their implications will be discussed in turn.

First, the conclusion that LSTFCoDel has lower delay than CoDel was expected. CoDel employs a FIFO queue. When congestion occurs, packets become increasingly delayed at the end of the queue (until they are eventually dropped, or congestion relieves). This is a major drawback to using a FIFO queue. delay is not equitably shared across packets, those in back receive the most delay.

Swapping the FIFO queue for a priority queue and utilizing an appropriate priority function allows LSTFCoDel to equitably disperse delay across packets. When congestion is low or none, the average slack time will eventually level out and dequeing will occur the same as a FIFO queue. As congestion builds, the priority level will increase in turn. Arriving packets will have increased priority over packets already in the queue (note this behavior emerges as a result of slack time). This is in stark contrast to the FIFO queue used by CoDel.

The second and third points presented above can be discussed together, as one depends on the other. Variance and standard deviation are tied to the stability of the backing queue employed by the queuing discipline. CoDel employs a FIFO queue which are inherently stable queues. Delay does not jump around because enqueues always occur at the end of the queue and dequeues always occur at the front of the queue. 

On the other hand, priority queues are much less stable. Assuming enqueues still occur at the end of the queue (as seen in LSTFCoDels implementation), dequeues jump around the queue. Priority queues are really only stable if the priority function produces stable values. If each successive iteration of the function produces an output of high variance to the prior output, then the queue will become unstable. In terms of LSTFCoDel, as long as congestion does not occur, the priority function will produce stable values leading to a stable queue.

``Table~\ref{table:lstfslack}'' shows the mean, variance, and standard deviation of the slack time for each value of $\alpha$ tested over the course of this research. Similar to the statistics presented in ``Table~\ref{table:lstfdelay}'', the variance and standard deviation of the slack time is quite large. This is to be expected given that slack time attempts to predict expected queuing delay.

\begin{table}[htbp]
\caption{Slack Statistics for LSTFCoDel $\alpha$ Levels}
\begin{center}
\begin{tabular}{|c|c|c|c|}
\hline
     Forgetfulness ($\alpha$) & Avg. (s) & Var. (s) & Std. Dev. (s) \\
     \hline
     
     \hline
     0.125 & 0.00703307 & 0.000136175 & 0.0116694 \\
     \hline
     0.250 & 0.00677347 & 0.000203010 & 0.0142482 \\
     \hline
     0.375 & 0.00895621 & 0.000665736 & 0.0258019 \\
     \hline
     0.500 & 0.00509139 & 0.000113650 & 0.0106607 \\
     \hline
     0.625 & 0.00770287 & 0.000433476 & 0.0208201 \\
     \hline
     0.750 & 0.00742405 & 0.000457980 & 0.0214005 \\
     \hline
     0.875 & 0.00816796 & 0.000540324 & 0.0232449 \\
     \hline
\end{tabular}
\label{table:lstfslack}
\end{center}
\end{table}

However, much like with ``Table~\ref{table:lstfdelay}'', ``Table~\ref{table:lstfslack}'' does not provide me with any additional information on deciding which $\alpha$ value is best for LSTFCoDel. However, the $\alpha$ value of 0.5 does coincide with having the lowest mean of all $alpha$ values, just like in ``Table~\ref{table:lstfdelay}''. These results indicate that slack time is at least approximating actual delay in the right direction, although it is not exact.

\subsection{Comparing CoDel and LSTFCoDel}
This section will provide evidence to support the hypothesis that LSTFCoDel produces shorter average queuing delay than CoDel. Because the data set for this experiment was too large (billions of data points) to load into memory on my machine, I instead made the decision to generate sample data in R \cite{r_manual} using the population data for CoDel as well as the population data for LSTFCoDel (where $\alpha$=0.5). Statistical tests were computed using the Basic Statistics and Data Analysis package (BSDA) \cite{r_bsda_manual}. The distribution was assumed to be normal on the basis of the Central Limit Theorem. Finally, the sample size was $n = 500$ for both the CoDel sample and LSTFCoDel sample.

As both samples are normally distributed, this allowed for the use of the t-test. The t-test presented here was conducted under the following parameters:
\begin{enumerate}
    \item The t-test is a two-sample t-test for a difference between two means with unequal variances (Welch t-test).
    \item The sample size, n is 500 for both samples.
    \item $\mu_1$ is the mean of the CoDel sample and $\mu_2$ is the mean of the LSTFCoDel sample.
    \item The null hypothesis is that $\mu_1 - \mu_2 = 0$.
    \item The alternative hypothesis is that $\mu_1 - \mu_2 > 0$.
    \item The t-test was conducted with an $\alpha$ value of 0.05.
\end{enumerate}

\begin{table}[htbp]
\caption{One-Sided Welsh T-Test for CoDel and LSTFCoDel (Forgetfulness = 0.5)}
\begin{center}
\begin{tabular}{|c|c|}
\hline
     Statistic & Value \\
     \hline
     
     \hline
     T-Statistic & 32.774 \\
     \hline
     Degrees of Freedom & 513.92 \\
     \hline
     P-Value & $< 2.2*10^{-16}$ \\
     \hline
     95\% CI Lower Bound & 0.02595785 \\
     \hline
     95\% CI Upper Bound & $\infty$ \\
     \hline
\end{tabular}
\label{table:delay_ttest}
\end{center}
\end{table}

Under normal conditions, the results displayed in ``Table~\ref{table:delay_ttest}''. would be considered highly abnormal by most statisticians. However, I refer the reader to ``Table~\ref{table:codel_delay_metrics}'' and ``Table~\ref{table:lstfdelay}''. The difference between the average mean for CoDel and LSTFCoDel is approximately 70\% for each forgetfulness value. An obvious issue that arises here is the difference in variance and standard deviation between the two algorithms. This difference fundamentally arises as a result of CoDel employing a FIFO queue and LSTFCoDel employing a priority queue (discussed thoroughly in earlier sections). 

\begin{table}[htbp]
\caption{F-Test for Comparison of Two Variances for CoDel and LSTFCoDel (Forgetfulness = 0.5)}
\begin{center}
\begin{tabular}{|c|c|}
\hline
     Statistic & Value \\
     \hline
     
     \hline
     F-Statistic & 0.014954 \\
     \hline
     Numerator DoF & 499 \\
     \hline
     Denominator DoF & 499 \\
     \hline
     P-Value & $< 2.2*10^{-16}$ \\
     \hline
     95\% CI Lower Bound & 0.01254431 \\
     \hline
     95\% CI Upper Bound & 0.01782548 \\
     \hline
\end{tabular}
\label{table:delay:ftest}
\end{center}
\end{table}

The reader may find themselves asking, "Why didn't he perform statistical tests on queue length or slack time?" The answer is quite simple. First, I am only concerned with comparing the average delay over time of CoDel and LSTFCoDel. The queue length is provided to further elaborate on the situation that is happening at both queues. Second, slack time does not exist in CoDel, thus I have nothing to compare against with CoDel. Astute observers would note that I do not perform any correlation testing between delay and slack time. This is extremely important and could lead to further insight into which $\alpha$ value is best for LSTFCoDel. However, this particular aspect is left to future work with a smaller data set.

\section{Drawbacks and Future Work}
LSTFCoDel as it is described here does possess some limitations. First and foremost is that it is not a stable algorithm. That is, LSTFCoDel's delay is spread much wider and possesses a far higher variance than CoDel. However, as mentioned, the implementation of LSTFCoDel within NS-2 does not utilize a true priority queue, rather it maintains CoDels FIFO queue and uses a shadow queue in the form of a standard template library multimap. The variance in delay experienced within by LSTFCoDel presented here is likely due to the time trade-offs needed in the accounting process between the two queues. For instance, in this implementation of LSTFCoDel, rather than constant dequeuing on the order of O(1) (as seen in FIFO), LSTFCoDel must deque linearly on the order of O(N) (the algorithm must search the FIFO queue from the beginning for the corresponding packet pulled from the multimap), incurring a significant penalty. Readers should be aware of this when analyzing the results for themselves. Future implementations should use a true priority queue for LSTFCoDel.

There is at least one other possible solution to fixing LSTFCoDel's variance problem. Instead of constantly servicing the packet with least slack time, I could relax this constraint. I could allow LSTFCoDel to borrow the three stage AQM mechanism from RED \cite{red_paper}. That is, in the first stage, act as LSTF, servicing packets with least slack time. In the second stage, service the least slack time packet with some probability, otherwise service the packet with most slack time. Finally, in the last stage, service the packet with the most slack time. The motivation here is similar to the motivation for LSTFCoDel. As the queue builds in size, it is guaranteed that some packets will be delayed more than others. It is entirely possible that LSTFCoDel is simply swapping the most heavily delayed packets from those that arrive with congestion against those already in queue when congestion first occurs (because ``Eq.~\eqref{slack_calc}'' and ``Eq.~\eqref{classify_func}'' cause heavily congested packets to be serviced first).

Of course, this modified version of LSTFCoDel would need some minimum and maximum threshold to determine which stage the algorithm is operating in. However, similar to RED, the transition between stages could be actualized using the average queuing delay. The only issue with this approach is that now, average queuing delay is used in two related calculations: the priority function, and threshold management. This particular improvement to LSTFCoDel is currently being investigated by the author.

Recall the enqueuing, dequeuing, and dropping state of LSTFCoDel as they are dramatically different from CoDel. When a packet arrives at an LSTFCoDel switching device, a slack time is determined for the packet based on the slack function presented in ``Eq.~\eqref{slack_calc}''. The inverse of one plus this slack time is then assigned as the packets priority in the queue as determined by ``Eq.~\eqref{classify_func}''. Also recall that the primary determining factors of slack time are historical and observed delay. These factors all slide around as the queue becomes congested and decongested. However, once a priority value is assigned to a packet it does not change. Perhaps LSTFCoDel should update the priority values of packets in the queue with each iteration of the algorithm? A definite drawback to this approach is added overhead in maintaining the priority invariant. If we change the priority of every packet in the queue each time a packet is enqueued, then we must also reorganize the queue or priority mechanism to maintain order. This introduces additional overhead into the algorithm.

With LSTFCoDel's dequeue operation, a packet is selected from the queue with the least slack value (in accordance with LSTF). This process holds true until the underlying CoDel control loop detects bad queue and enters into the packet dropping state. As stated in \cite{codel_rfc}, in packet dropping state, CoDel should only drop one packet then exit the state. However, excessive delay will keep CoDel in packet dropping state until all sources of congestion relieve their sending rates (if possible, transport protocols like UDP do not care about congestion). The dequeue operation takes on a different form when CoDel drops a packet. Instead of dequeuing from the beginning of the queue (that is, dequeuing the least slack time packet), LSTFCoDel will drop the packet with the highest slack time. Additional testing on this mechanism is needed, as I have not explicitly tested it with this study.

The last major drawback with LSTFCoDel presented in this paper is that it has only been tested in simulation. To quote Dr. Ye Zhu of Cleveland State University, "Its the Internet, anything can happen!" Testing done in a controlled environment can only reveal so much. Real world implementation and testing is needed to validate the intricacies of LSTFCoDel.

\section{Conclusion}

LSTFCoDel is a versatile algorithm capable of handling packets both at times of no congestion as well as peak congestion. Based on the principles of LSTF and building on top of the CoDel AQM algorithm, LSTFCoDel is able to provide priority service to packets that need it most. The magic of LSTFCoDel is that it transparently lives on top of CoDel. No underlying modifications of CoDel's control loop logic are necessary to support the algorithm. Implementations merely have to provide the slack and priority functions described in this paper and swap CoDels backing queue with a priority queue. 

I have posed an argument in favor of LSTFCoDel and its necessity. Additionally, I have provided overwhelming statistical evidence to the efficacy of LSTFCoDel. Much of this work builds upon the work outline in \cite{codel_rfc} and \cite{lstf_paper}. The intention of this algorithm is not to replace CoDel. Rather, the work presented here is intended as an improvement over CoDel meant to decrease average queuing delay and address service unfairness when congestion occurs. Outside of congestion LSTFCoDel should perform just like FIFO (once the average slack time $\gamma$ settles). However, it should be noted that until the variance issues are resolved, LSTFCoDel is inherently less stable than CoDel (I have provided reasons why that might be as well as ways to address them).

\bibliographystyle{IEEEtran}
\bibliography{refs}

\end{document}